\documentclass[onecolumn]{elsart3p} 
\usepackage{ifpdf}

\usepackage{graphicx}
\usepackage{amssymb}
\usepackage{amsmath}
\usepackage{dcolumn}
\usepackage{bm}

\begin{document}

\begin{frontmatter}

\title{Effect of the shape of periodic forces and second periodic forces on
horseshoe chaos in Duffing oscillator}
\author[kgsarts]{V. Ravichandran},
\author[kgsarts]{V. Chinnathambi},
\author[bdu]{S. Rajasekar\corauthref{cor1}},
\author[sing]{Choy Heng Lai}

\corauth[cor1]{Corresponding author.  E-mail: srj.bdu@gmail.com; Fax +91 
431 2407093; Phone +91 431 2407057}

\address[kgsarts]{Department of Physics, Sri K.G.S. Arts College, %
Srivaikuntam 628 619, Tamilnadu, India}
\address[bdu]{School of Physics, Bharathidasan University, %
Tiruchirappalli 620 024, Tamil Nadu, India}
\address[sing]{Department of Physics, National University of Singapore, %
Singapore 117542}

\begin{abstract}
The effect of the shape of six different periodic forces and second periodic forces on the onset of horseshoe chaos are studied both analytically and numerically in a Duffing oscillator. The external periodic forces considered are sine wave, square wave, symmetric saw-tooth wave, asymmetric saw-tooth wave, rectified sine wave, and modulus of sine wave. An analytical threshold condition for the onset of horseshoe chaos is obtained in the Duffing oscillator driven by various periodic forces using the Melnikov method. Melnikov threshold curve is drawn in a parameter space. For all the forces except modulus of sine wave, the onset of cross-well asymptotic chaos is observed just above the Melnikov threshold curve  for onset of horseshoe chaos. For the modulus of sine wave long time transient motion followed by a periodic attractor is realized. The possibility of controlling of horseshoe and asymptotic chaos in the Duffing oscillator by an addition of second periodic force is then analyzed. Parametric regimes where
suppression of horseshoe chaos occurs are predicted. Analytical prediction is demonstrated through direct numerical simulations. Starting from asymptotic chaos we show the recovery of periodic motion for a range of values of amplitude and phase of the second periodic force. Interestingly, suppression of chaos is found in the parametric regimes where the Melnikov function does not change
sign.
\end{abstract}

\end{frontmatter}

\maketitle

\section{Introduction}

The dynamics of nonlinear oscillators and circuits are very often studied with  external force being of the form $f\sin\omega t$ or $f\cos\omega t$. Other forms of forces such as square wave, rectified sine wave, rectangular wave, etc. can be generated and applied to dynamical systems. The study of the effect of such periodic forces will be helpful to choose a suitable external drive in creating and controlling nonlinear behaviours. In recent years there are reports on the effect of different forces on certain nonlinear phenomena \cite{Chacon05,Konishi03,Zengrong04,Ge04,Lai04,VMG07}. Analysis of features of a particular dynamics with various periodic forces and a detailed comparative study of effects induced by them will be of great use. It is also important to explore the utility and  applicability of analytical methods such as multiple-scale perturbation theory and Melnikov method to the
systems driven by periodic forces other than $f\sin\omega t$ and
$f\cos\omega t$.

Motivated by the above, in the present paper we wish to study the occurrence of horseshoe chaos in Duffing oscillator driven by different periodic forces applying Melnikov analytical method. Recently, this method has been  applied to study the Duffing and Duffing-van der Pol oscillators with different external perturbations \cite{Nana05,Tang06,RWang06,Balibrea05,Wang06,Yang06,Jing06,Gao05}. The Melnikov method has also been applied to other nonlinear
systems. 

In the present paper, we consider the perturbed Duffing double-well oscillator 
%
\begin{subequations}
\begin{eqnarray}
 \dot{x} &=&  y \;,\\
 \dot{y} &=&  \omega_{0}^{2} x - \beta x^{3} + \epsilon \left[
                - \alpha y+ F(t) \right]\;,
\end{eqnarray}
\label{eq1}
\end{subequations}
%
where$\;\alpha$ is the damping constant,$\;\omega_{0}^{2}$ is the
natural frequency,$\;\beta$ is the stiffness constant which plays
the role of nonlinear parameter, $F(t)$ is an external periodic
force, $\epsilon$ is a small parameter, $\omega_{0}^{2} >$ 0, $\beta >$ 0.
We study the occurrence of horseshoe chaos with different
periodic forces such as sine wave, square wave, symmetric
saw-tooth wave, asymmetric saw-tooth wave, rectified
sine wave and modulus of sine wave. In section 2, we obtain the Melnikov threshold condition for the transverse intersection of homoclinic orbits for the system (\ref{eq1})
separately for each of the above periodic forces. In
section 3, we plot the Melnikov threshold curve in the
($f-\omega$) parameter space for all the forces where $f$ and
$\omega$ are the amplitude and frequency of the external periodic
force. We verify the analytical prediction with the numerically
calculated critical values of $f$ at which transverse
intersections of stable and unstable manifolds of the saddle
occur. The Melnikov threshold value is also compared with the
onset of asymptotic chaos wherever possible. Only for the modulus
of sine wave long time transient motion instead of asymptotic
chaos near the Melnikov threshold curve is found. We characterize
the transient dynamics using the average transient time. In
section 4, we consider the system (\ref{eq1}) driven by two sine and
modulus of sine forces and analyze the effect of the amplitude
and phase of the second force. Starting from horseshoe chaos and
asymptotic chaos we show the possibility of suppressing them. The
control regions can be identified by the Melnikov method. Finally, we
end up with conclusion in section 5.

\section{Calculation of Melnikov function}

The unperturbed part of the system (\ref{eq1}) with $\epsilon$=0 has one saddle point
($x^{*},y^{*}$) = (0,0), and two center type fixed points,
($x^{*},y^{*}$) =($\pm \sqrt{\omega_{0}^{2}/\beta},0)$. The two homoclinic orbits connecting the saddle to itself are given by
\begin{eqnarray}
  W^{\pm}(x_{h}(t),\;
   y_{h}(t))& = & \left(\pm\sqrt{2\omega_{0}^{2}/\beta}
                \; \mbox{sech} \left(\sqrt{\omega_{0}^{2}}\;
                t\right), \right. \nonumber \\ 
           & & \quad \left. \mp  \sqrt{2/\beta}\; \omega_{0}^{2}
                    \; \mbox{sech} \left(\sqrt{\omega_{0}^{2}}\;
                t \right) \tanh \left(\sqrt{\omega_{0}^{2}}\; t
                \right)\right).
\label{eq2}
\end{eqnarray}
The Melnikov function $M(t_{0})$ measures the distance between the stable manifold $(W_{\mbox{s}})$ and unstable manifold $(W_{\mbox{u}})$ of a saddle. When the two orbits are
always separated then the sign of $M(t_{0})$ always remains same. $M(t_{0})$ oscillates
when the orbits $W_{\mbox{u}}$ and $W_{\mbox{s}}$ intersect transversely (horseshoe dynamics). A zero of $M(t_{0})$  corresponds to a tangential intersection. The occurrence of
transverse intersections implies that the Poincar\'e map of
the system has the so-called horseshoe chaos \cite{Chacon05,Guckenheimer90,Wiggins90}. 

For the Duffing eq.(\ref{eq1}), the Melnikov function is
\begin{equation}
  M(t_{0}) = \int\limits_{\,\,\,\,-\infty}^{\,\,\,\,\infty} y_{h}
             \left[-\alpha \;y_{h} +  F(\tau + t_{0})\right] d\tau.
\label{eq3}
\end{equation}
In the following we calculate the Melnikov function for
the system (\ref{eq1}) with different periodic forces.

For the system (\ref{eq1}) driven by the force $F(t)=f_{\sin} \sin \omega t$, the Melnikov
integral (\ref{eq3}) is worked out as
\begin{subequations}
\begin{eqnarray}
    M_{\sin}^{\pm}(t_{0}) = A \pm f_{\sin} B \cos\omega t_{0},
\label{eq4a}
\end{eqnarray}
where
\begin{eqnarray}
  A = -\frac{4\;\alpha}{3 \beta}\;(\omega_{0}^{2})^{3/2}\;,\;\;\;\;
        B = \sqrt{2/\beta}\; \pi \omega \;
        \mbox{sech}\left(\frac{\pi\omega}{2\sqrt{\omega_{0}^{2}}} \;\right).
\label{eq4b}
\end{eqnarray}
\end{subequations}
\noindent For the square wave $ F(t) = F_{\mbox{sq}}(t+{2\pi}/{\omega})
        = f_{\mbox{sq}}\;\mbox{sgn}(\sin \omega t)$ where $\mbox{sgn}$${(y)}$ is sign of $y$, using its Fourier series we obtain
\begin{subequations}
\begin{eqnarray}
 M_{\mbox{sq}}^{\pm}(t_{0}) = A \;\pm\; f_{\mbox{sq}}\sum_{n=1}^{\infty}
                         B_{n}\cos(2n-1)\omega t_{0},
\label{eq5a}
\end{eqnarray}
where
\begin{eqnarray}
  B_{n} = \frac{4\sqrt{2}\;\omega}{\sqrt{\beta}}\; \mbox{sech}
           \left(\frac{(2n-1)\pi\omega}
             {2\sqrt{\omega_{0}^{2}}}\right).
\label{eq5b}
\end{eqnarray}
\end{subequations}
For the symmetric saw-tooth wave of the form
\begin{equation}
   F_{\mbox{sst}}(t)=F_{\mbox{sst}}\left(t+{2\pi}/{\omega}\right)=
               \left\{
\begin{array}{ll}
  {{4f_{\mbox{sst}}t}/{T}}, & \hskip 10pt 0 \leq t < {\pi}/{2\omega}\\
    {{-4f_{\mbox{sst}}t}/{T}} + 2f_{\mbox{sst}}, & \hskip 10pt
    {\pi}/{2\omega} \leq t <
    {3\pi}/{2\omega}\\
   {{4f_{\mbox{sst}}t}/{T}}- 4f_{\mbox{sst}},  & \hskip 10pt
   {3\pi}/{2\omega} \leq t <
   {2\pi}/{\omega}\\
\end{array}
\right.
\label{eq6}
\end{equation}
the Melnikov function is
\begin{subequations}
\begin{eqnarray}
   M_{\mbox{sst}}^{\pm}(t_{0}) = A\; \pm
    \;f_{\mbox{sst}}\sum_{n=1}^{\infty}\; B_{n}\cos(2n-1)\omega t_{0},
\label{eq7a}
\end{eqnarray}
where
\begin{eqnarray}
  B_{n} =
   \frac{8\sqrt{2}\omega}{\pi\sqrt{\beta}}\;\frac{(-1)^{n+1}}{(2n-1)}
   \;\mbox{sech}\left[\frac{(2n-1)\pi\omega}{2\sqrt{\omega_{0}^{2}}}\right].
\label{eq7b}
\end{eqnarray}
\end{subequations}
For the asymmetric saw-tooth wave 
\begin{subequations}
\begin{eqnarray}
   F_{\mbox{ast}}(t) =
   F_{\mbox{ast}}\left(t+{2\pi}/{\omega}\right)=\left\{
\begin{array}{ll}
  {2f_{\mbox{ast}}t}/{T}, & \hskip 10pt 0 \leq t < {\pi}/{\omega}\\
  {{2f_{\mbox{ast}}t}/{T}}-2f_{\mbox{ast}}, & \hskip 10pt
  {\pi}/{\omega} \leq t <
  {2\pi}/{\omega},\\
\end{array}
 \right. \label{eq8a}
\end{eqnarray}
we obtain
\begin{eqnarray}
  M_{\mbox{ast}}^{\pm}(t_{0}) = A \;\pm
    f_{\mbox{ast}}\sum_{n=1}^{\infty}\; B_{n}\cos n \omega t_{0}, \quad B_{n}=\frac{2\sqrt{2}\omega}{\sqrt{\beta}}\;(-1)^{n+1}
         \;\mbox{sech}\left(\frac{n\pi\omega}
         {2\sqrt{\omega_{0}^{2}}}\right).
\label{eq8b}
\end{eqnarray}
\end{subequations}
For the rectified sine wave
\begin{equation}
F_{\mbox{rec}}(t) =
   F_{\mbox{rec}}\left(t+{2\pi}/{\omega}\right)=\left\{\begin{array}{ll}
   f_{\mbox{rec}} \sin\omega t, & \hskip 10pt 0 \leq t<{\pi}/{\omega}  \\
   0, & \hskip 10pt  {\pi}/{\omega} \leq t < {2\pi}/{\omega}\\
\end{array}\right.
\label{eq9}
\end{equation}
we find
\begin{subequations}
\begin{eqnarray}
M_{\mbox{rec}}^\pm(t_{0}) = A\;\pm\; 2
f_{\mbox{rec}}\sum_{n=1}^{\infty}\; B_{n}\;C_{n}\sin n\omega
t_{0},
\label{eq10a}
\end{eqnarray}
where
\begin{eqnarray}
  B_n = \frac{\sqrt{8}\omega\omega_{0}^{2}}{\pi \sqrt{\beta}} \;
         \frac{1}{\omega^{2}-n^{2}}, \quad 
  C_{n} =
\int\limits_{\;\;\;\;(2n-2)\pi/{\omega}}^{\;\;\;\;(2n-1)\pi/\omega}
\;\mbox{sech}\left(\sqrt{\omega_{0}^{2}}\tau\right) \;
\mbox{tanh}\left(\sqrt{\omega_{0}^{2}}\tau\right) \; \sin
n\omega\tau \mathrm{d} \tau.
\label{eq10b}
\end{eqnarray}
\end{subequations}
\noindent For the modulus of sine wave $ F_{\mbox{msw}}(t) = F_{\mbox{msw}}(t+ {2\pi}/{\omega}) = f_{\mbox{msw}}\;\sin \left(\omega t/2\right)$, we find
\begin{subequations}
\label{eq11}
\begin{eqnarray}
  M_{\mbox{msw}}^{\pm} (t_0)  
       & = &  A \pm f_{{\mbox{msw}}} \sum_{n=1}^{\infty} B_n  \sin n \omega t_0 \;,  \\
   B_n & = &  \frac{8 \sqrt{2} \, \omega \, n^2}{(4n^2-1) \sqrt{\beta}}  \,
               {\mbox{sech}} \left( \frac{n \pi \omega}{2 \sqrt{\omega_0^2}} \right)
                \;. 
\end{eqnarray}
\end{subequations}

\section{Horseshoe chaos and strange attractor}

In this section, we compute the Melnikov threshold values for
horseshoe chaos and compare it with numerical prediction. For the
 sine wave force the threshold
condition for transverse intersections of stable manifolds
$(W_{\mbox{s}}^{\pm})$  and unstable manifolds
$(W_{\mbox{u}}^{\pm})$ is
\begin{equation}
|f_{\sin}|\;\geq\; |f_{M}| =
\frac{2\sqrt{2}\alpha(\omega_{0}^{2})^{{3}/{2}}}{3\pi\omega
\sqrt{\beta}
}\;\mbox{cosh}\left(\frac{{\pi\omega}}{{2\sqrt{\omega_{0}^{2}}}}\right) \;.
\label{eq12}
\end{equation}
%
%
%
%
For the other forces $M(t_{0})$ is a convergent series. In these cases
the threshold values of $f$ can be determined numerically. For the
rectified sine wave force $C_{n}$'s in eq.(\ref{eq10b}) have to be
calculated numerically. Throughout our study we fix the parameters
in eq.(\ref{eq1}) as $\alpha = 0.5$, $\beta = 1$, $\omega_{0}^{2} = 1$
and $\omega = 1$. $f_{M}$
can be calculated numerically as follows. For the forces other
than sine and modulus of sine wave, we cannot write the sufficient
condition for the existence of simple zeros of $M(t_{0})$.
Therefore, we identify the occurrence of homoclinic bifurcations
by numerically measuring the time $\tau_{M}$ elapsed between two
successive zeros of $M(t_{0})$. $\tau_{M}$ is calculated for a range of amplitude of the
external force. The value of $f$ at which first time $M(t_{0})$
changes sign and thereby giving finite $\tau_{M}$ is the Melnikov
threshold value for homoclinic bifurcation.

\begin{figure}
\begin{center}
\includegraphics[width=0.6\columnwidth,clip]{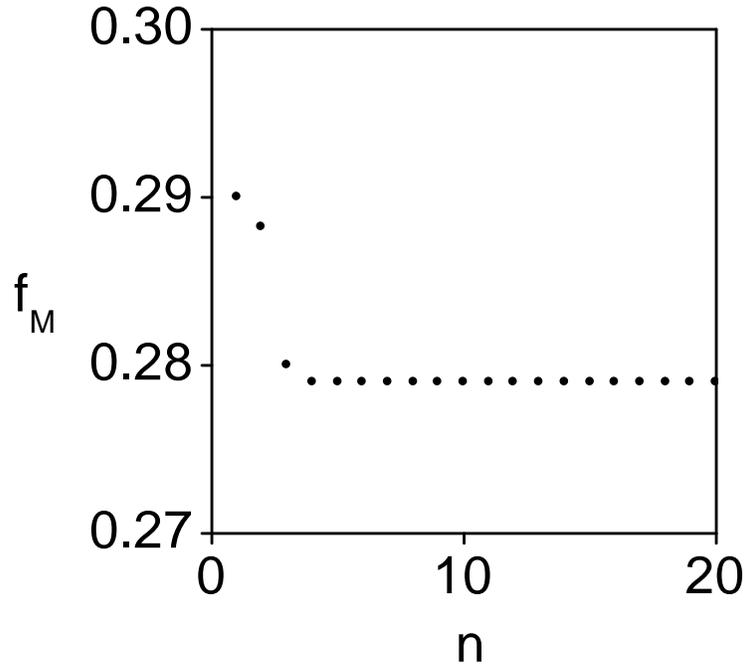}
\end{center}
\caption{$f_M$ versus $n$, the number of terms in the summation in
eq.(\ref{eq5a}), for $\alpha$ = 0.5, $\beta$ = 1, $\omega_{0}^{2}$ = 1
and $\omega$ = 1 when the system (\ref{eq1}) is driven by the square wave
force. Variation in $f_{M}$ converges to a constant value with
increase in $n$.} \label{Fig1}
\end{figure}

\begin{figure}
\begin{center}
\includegraphics[width=0.5\columnwidth,clip]{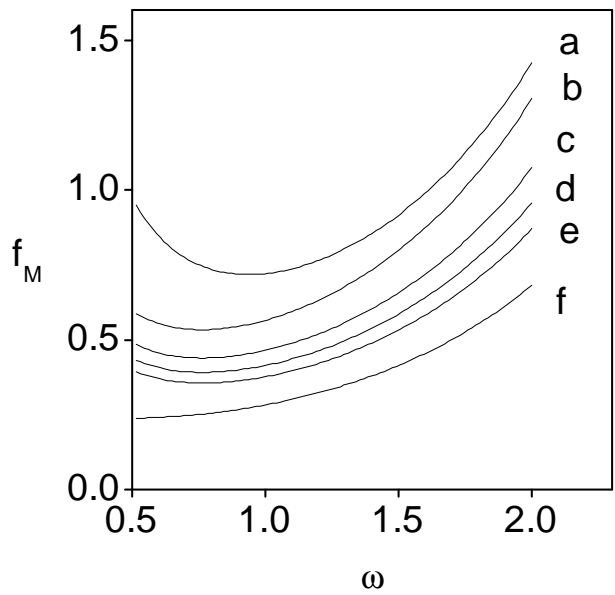}
\end{center}
\caption{Melnikov threshold curves for horseshoe chaos in the $(f
- \omega)$ plane for the system (\ref{eq1}) driven by the forces (a)
asymmetric saw-tooth wave, (b) rectified sine wave, (c) symmetric
saw-tooth wave, (d) modulus of sine wave, (e) sine wave and (f)
square wave. The values of the parameters in eq.(\ref{eq1}) are $\alpha$ =
0.5, $\beta$ = 1 and $\omega_{0}^{2}$ = 1.}\label{Fig2}
\end{figure}

Figure (\ref{Fig1}) shows the plot of $f_{M}$ versus $n$, the number of
terms in the summation, eq.(\ref{eq5a}), for the square wave force. $f_{M}$
converges to a constant value with increase in $n$. For $n > $ 10,
the variation in $f_{M}$ is negligible. Similar result is found
for the other forces also. In our numerical calculation of $f_{M}$
we fix $n$ = 50. Figure (\ref{Fig2}) shows the plot of the threshold curves for horseshoe
chaos in the $(f-\omega)$ parameter plane. Below the threshold
curve no transverse intersection of stable and unstable manifolds
of the saddle occurs and above the threshold curve the transverse
intersection of orbits of the saddle occurs. The threshold curves
are nonintersecting. Smooth variation of $f_{M}$ is found for all
the forces considered in our study. The variation of $f_{M}$ with
$\omega$ is similar for all the periodic forces. Among the six
forces, $f_{M}$ is maximum for the asymmetric saw-tooth wave and
is minimum for the square wave. Thus, onset of horseshoe chaos can
be either delayed or advanced by an appropriate choice of periodic
force.

We verify the analytical prediction by numerically computing the
stable and unstable manifolds of the saddle. In fig.(\ref{Fig3}) we plotted the
orbits of the saddle for two values of $f$ $-$ one for $f < f_{M}$
and another for $f > f_{M}$ for each of the forces. For clarity
only part of the manifolds are shown. In the left side subplots,
for $f=0.2$, the stable and unstable orbits are well separated. In
the right side subplots, for $f > f_{M}$, we can clearly notice
transverse intersections of orbits at certain places. The critical values of $f_{M}$ and $f_{\mbox{cwc}}$ (at which onset of cross-well chaos occurs) for various forces are given in
Table 1. The numerical result agrees well with the theoretical
prediction.
\begin{table}[!ht]
\caption{Critical values of $f_{M}$ and $f_{\mbox{cwc}}$ (the
values of $f$ at which cross-well chaos
 occur) for the Duffing eq.(\ref{eq1}) for various forces with
$\alpha = 0.5$, $\omega_{0}^{2} = 1.0$, $\beta = 1.0$ and $\omega
= 1.0$.}
\centering
\begin{tabular}{lllll}
   \hline
   \noalign{\smallskip}
No. & Force & $f_M$ & & $f_{\mbox{cwc}}$ \\
 \noalign{\smallskip}
   \hline
   \noalign{\smallskip}
 1. & Sine wave & 0.3565 & &0.3833 \\
 2. & Square wave & 0.2745 & & 0.2910 \\
 3. & Symmetric saw-tooth wave & 0.4629 & & 0.4792 \\
 4. & Asymmetric saw-tooth wave & 0.7155 & & 0.7326 \\
 5. & Rectified sine wave & 0.5026 & & 0.5192 \\
 6. & Modulus of sine wave & 0.3985 & & $---$ \\
   \noalign{\smallskip}
   \hline
\end{tabular}
\end{table}

For the system (\ref{eq1}) driven by modulus of sine wave onset of
cross-well chaos (asymptotic) is not observed near the Melnikov
threshold value. However, transient motion followed by a long time
periodic motion and hysteresis phenomenon are observed. Different paths are followed when $f$ is increased from $0$ to $0.5$ and decreased from $0.5$ to $0$. Onset of jumping motion from left-well to right-well is found to occur at $f_{\mbox {h}} =0.440547$. Long time transient
motion is found for $f$ values just below $f_{\mbox {h}}$. For a value of $f$, we
calculated the transient time $\tau_{T}$, the time taken to settle
on the right-well attractor for a set of 900 initial conditions in
the region $x\in[-0.9, -0.6]$, $\dot{x}\in [-0.2, 0]$. Then,
average transient time $\tau$ is calculated for a range of values
of $f$. Figure (\ref{Fig4}) shows the variation of $\tau$ as a function of
$(f - f_c)$. $\tau$ is found to scale as $1.04166 (f -
f_c)^{-0.49}$. 

\section{Effect of second periodic forces}

In this section, we consider the problem of controlling of chaos by the addition of
second weak periodic forces. We consider only
the two forces, namely, sine wave and modulus of sine wave. The
effect of other forces can also be studied.

In eq.(\ref{eq1}) we choose $F(t)=f_{\sin} \sin \omega t + g_{\sin} \sin(\Omega t + \phi)$ 
where $\phi$ is the phase shift. Recently, Zambrano et al \cite{Zambrano06}
studied suppression of chaos by the phase difference between the
two external periodic forces. Wang et al \cite{Wang06} found the occurrence
of strange non-chaotic attractor in the Duffing oscillator for a
range of values of phase $\phi$ with the force $f \cos(\omega t) +
g \cos(\Omega t + \phi(t))$.

For the system (\ref{eq1}), the Melnikov function is
\begin{subequations}
\label{eq14}
\begin{eqnarray}
  M_{\mbox{sin}}^\pm (t_{0}) = A \pm B\; f_{\sin} \cos \omega t_{0} \pm C\;
g_{\sin} \cos ( \Omega t_{0} + \phi),
\label{eq14a}
\end{eqnarray}
where $A$ and $B$ are given by eq.(\ref{eq4b}) and
\begin{eqnarray}
C = \sqrt{2/\beta}\;\pi \Omega \;\mbox{sech}\left(\frac{\pi \Omega}
          {2\sqrt{\omega_{0}^{2}}}\right).\;\;\;\;\;\;
           \;\;\;\;\;\;\;\;\;\;\;\;\;\;\;\;\;\;\;\;\;\;\;\;\;
\end{eqnarray}
\label{eq14b}
\end{subequations}
The choice $\Omega=\omega$ and $\phi=0$ is trivial. For $\Omega \neq \omega$ a
Melnikov condition similar to eq.(\ref{eq12}) cannot be written for
arbitrary values of $\omega$ and $\Omega$. So, we numerically
compute the time $\tau_{M}$ and identify the parametric regime
where $1/\tau_{M} \approx 0$. In the absence of the second
periodic force the horseshoe chaos is found to occur for $f$=0.359
and $\omega$ = 1. We study the possibility of suppression of chaos
by the addition of the second periodic force. Figure (\ref{Fig5}) shows the plot
of $1/{\tau_{M}}$ against $\Omega$ for $g=0.1$ and $\phi=0$. In this figure 
$1/{\tau_{M}}$ is zero (that is, $\tau_{M}$ is infinity) only at $\Omega$ = 0.5, 1.5 and $\Omega > 2.0$. This  implies that horseshoe chaos does not occur for these three values of
$\Omega$. For other values of $\Omega$, both $M^{+}(t_{0})$ and
$M^{-}(t_{0})$ oscillate and hence $1/{\tau_{M}}$ are nonzero.
These results are verified numerically. 

Next, we study the effect of $\phi$. For $\omega = \Omega$ and
$\phi \neq 0$ (which gives $B = C$), the condition for $M(t_{0})$ to change sign is given by
\begin{equation}
g_{\sin}^{2} + 2 f_{\sin}\;g_{\sin}\;\cos \phi +
f_{\sin}^{2}-(A^{2}/B^{2})\;\geq\; 0.
\label{eq15}
\end{equation}
The equality sign corresponds to tangential intersections of the
orbits of saddle. We denote $g_{\sin}^{+}$ and $g_{\sin}^{-}$ are the two roots of the eq.(\ref{eq15}) with equality sign. 

Figure (\ref{Fig6}) shows the plot of $g_{\sin}$ versus $\phi$ for
$\omega=\Omega=1$, and $f=0.359$. In the regions
$\bf{a}$, $\bf{b}$ and $\bf{c}$ enclosed by the continuous and
dashed curves in fig.(\ref{Fig6}), the sign of $M(t_{0})$ remains same and
transverse intersection of orbits of the saddle does not occur. In
the remaining region $M(t_{0})$ oscillates and horseshoe chaos
occurs. The above result is also confirmed by direct numerical
simulation. We note that suppression of horseshoe chaos can
be achieved for a range of values of $\phi$, $\Omega$ and
$g_{\sin}$.  A striking result is that suppression of asymptotic
chaotic motion is also found in the regions $\bf{a}$, $\bf{b}$ and
$\bf{c}$. In fig.(\ref{Fig7}) bifurcation diagram and maximal Lyapunov
exponent ($\lambda$) as a function of $g_{\sin}$ are reported. For
$\phi=0.5$ the maximal Lyapunov exponent $\lambda$ is found to be
negative in the interval $g_{\sin} \in [- 0.6077, 0.0108]$
while the interval of $g$ in which suppression of horseshoe chaos
predicted by the Melnikov method is  $[- 0.6097, 0.0103]$.
 For $\phi=\pi$, suppression of asymptotic chaos is found in the
interval $g_{\sin} \in [- 0.017, 0.735]$. The Melnikov  method
predicted interval of $g$ for suppression of horseshoe chaos is
$[-0.015, 0.735]$. Similar results are observed for various values
of $\phi$ in the regions $\bf{a}$, $\bf{b}$ and $\bf{c}$.

For the Duffing oscillator driven by two modulus of sine wave
%
\begin{equation}
M_{\mbox{msw}}^\pm (t_0) = A \pm f_{\mbox{msw}} \sum_{n=1}^{\infty} B_n 
            \sin (n \omega t_0) \pm g_{\mbox{msw}} \sum_{n=1}^{\infty}
             D_n \sin ( n \Omega t_0 + \phi) \;,
\label{eq16}
\end{equation}
where $A$ and $B_n$ are given by the eqs.(\ref{eq4b}) and (\ref{eq11}) respectively and $D_n$ is same as $B_n$ except that $\omega$ in $B_n$ is now replaced by $\Omega$. 
We numerically compute the time $\tau_{M}$ and identify the parametric regions where $1/{\tau_{M}}\approx 0$. 
%
%
%
Figure (\ref{Fig8}) shows the plot of threshold values of $g_{\mbox{msw}}$ versus $\phi$ for $\omega=\Omega=1$, and $f_{\mbox{msw}}=0.5$. In
the regions $\bf{a}$, $\bf{b}$ and $\bf{c}$ enclosed by the
continuous and dashed curves in fig.(\ref{Fig8}), the sign of $M(t_{0})$
remains same and transverse intersection of orbits of the saddle
 does not occur. In the remaining region $M(t_{0})$ oscillates and
horseshoe chaos occurs. The above result is also verified by
numerical simulation. For example, fig.(\ref{Fig9}) shows the part of the
stable and unstable orbits in the Poincar\'e map for two values of
$g_{\mbox{msw}}$ chosen outside and inside the region $\bf{b}$ for
$\omega$ = $\Omega=1$, $\phi = \pi$ and $f = 0.5$. The threshold
values of $g_{\mbox{msw}}$ are $g_{\mbox{msw}}^{+}$ = 0.615 and $g_{\sin}^{-}$ = 0.393. When
$g_{\mbox{msw}}$ is varied from say $-0.7$ horseshoe chaos does
not occur in the interval $g_{\mbox{msw}}^{-} < g <
g_{\mbox{msw}}^{+}$. In fig.(\ref{Fig9}a) transverse intersections are
seen  for $g_{\mbox{msw}} = - 0.1$. For $g_{\mbox{msw}} = 0.5$
(lying in the interval [$g_{\mbox{msw}}^{-}$,
$g_{\mbox{msw}}^{+}$]) the stable and unstable orbits are well
separated (fig.\ref{Fig9}b). Suppression of horseshoe chaos can be achieved for a range of
values of $\phi$, $\Omega$ and $g_{\mbox{msw}}$. The bifurcation diagram shows
absence of cross-well chaos (asymptotic) near the Melnikov
threshold value. However, long transient motion followed by a
periodic motion and hysteresis phenomenon are observed.

\section{Conclusion}

In the present paper we have performed a numerical and analytical
studies of homoclinic bifurcation in the Duffing oscillator driven
by different periodic forces and addition of second periodic
forces. Applying the Melnikov analytical method we obtained the
threshold condition for onset of horseshoe chaos, that is,
transverse intersections of stable and unstable branches of
homoclinic orbits. For the modulus of sine wave instead of cross-well chaos a long time
transient motion followed by a period-$T$ solution is observed.
Analytical prediction of horseshoe chaos is found to be in good
agreement with numerical simulation for all the forces. Suppression of horseshoe chaos and asymptotic chaos are found for a range of amplitude of the second periodic force and phase shift between the two external forces. With the good agreement obtained between theoretical and numerical predictions we emphasize that the Melnikov analysis can be
successfully used to predict the onset of chaos in the presence of
weak periodic perturbations and also the effect of such
perturbations on regular and chaotic dynamics.

In the Duffing oscillator driven by each of the different periodic forces considered in our study transverse intersection of left-well homoclinic orbits and
transverse intersection of right-well homoclinic orbits occur at
same critical values of, say, the amplitude $f$ of the force. This
is because in the absence of driving force and damping the
potential is symmetric and the two homoclinic orbits connecting
saddle to itself satisfy the relation $W^{-} = - W^{+}$. It is of
interest to consider the situation where $W^{-} \neq \pm  W^{+}$.
In this case, it is possible to have onset horseshoe chaos (as
well as asymptotic chaos) in the two wells at different threshold
values. This will be analyzed in future.

\ack

The work reported here forms part of a Department of Science and
Technology, Government of India research project of SR.


%
%

\begin{figure}
\begin{center}
\includegraphics[width=0.9\columnwidth,clip]{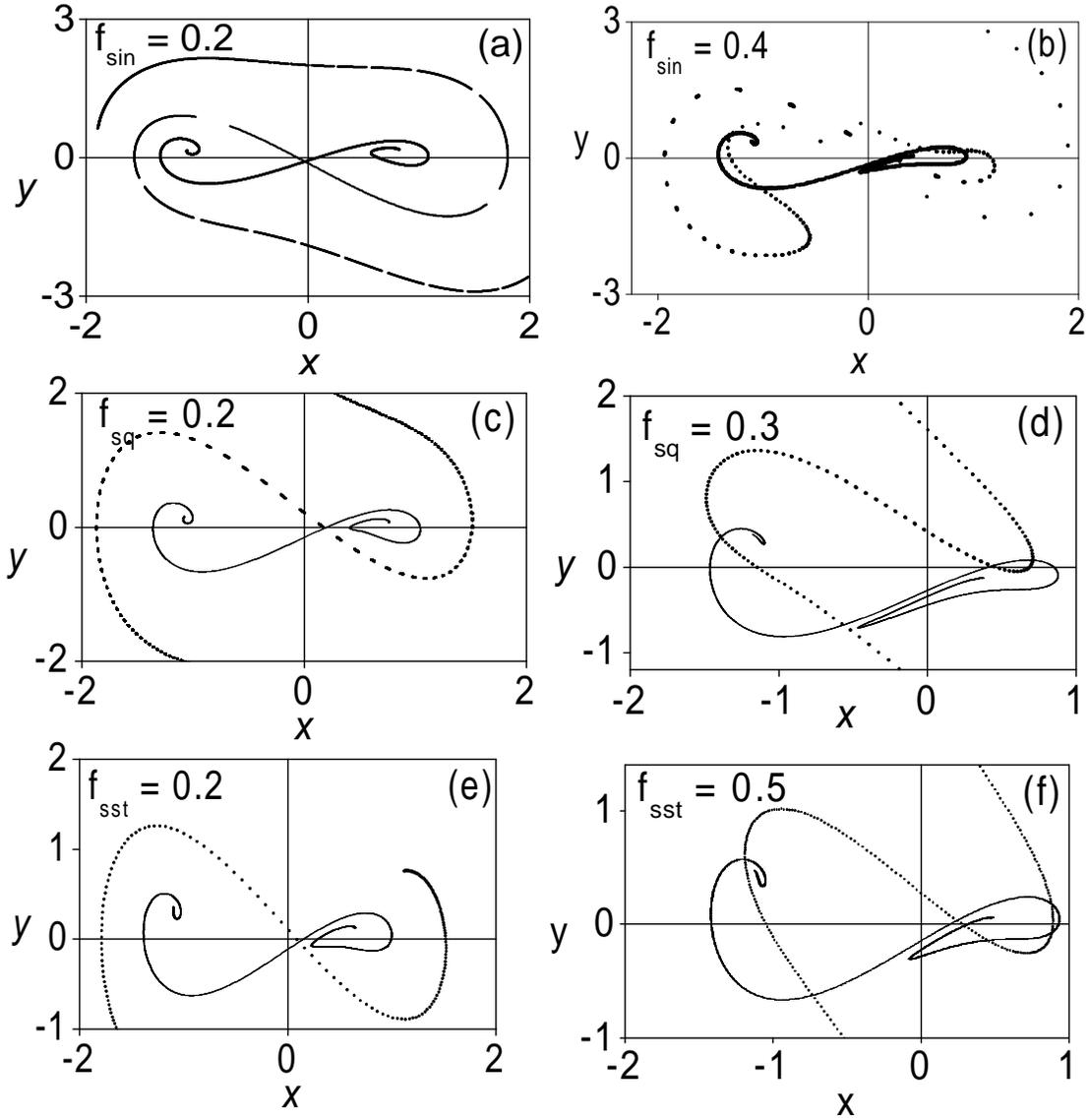}
\end{center}
\caption{Numerically computed stable and unstable manifolds of the
saddle fixed point of the system (\ref{eq1}). The system is driven by
(a-b) sine wave, (c-d) square wave, (e-f) symmetric saw-tooth
wave, (g-h) asymmetric saw-tooth wave, (i-j) rectified sine wave
and (k-l) modulus of sine wave. The other parameters are $\alpha$
= 0.5, $\beta$ = 1, $\omega_{0}^{2}$ = 1 and $\omega$ = 1. Left
side subplots are for $f < f_{M}$ while the right side subplots
are for $f > f_M.$}\label{Fig3}
\end{figure}
\addtocounter{figure}{-1}
\begin{figure}
\begin{center}
\includegraphics[width=0.9\columnwidth,clip]{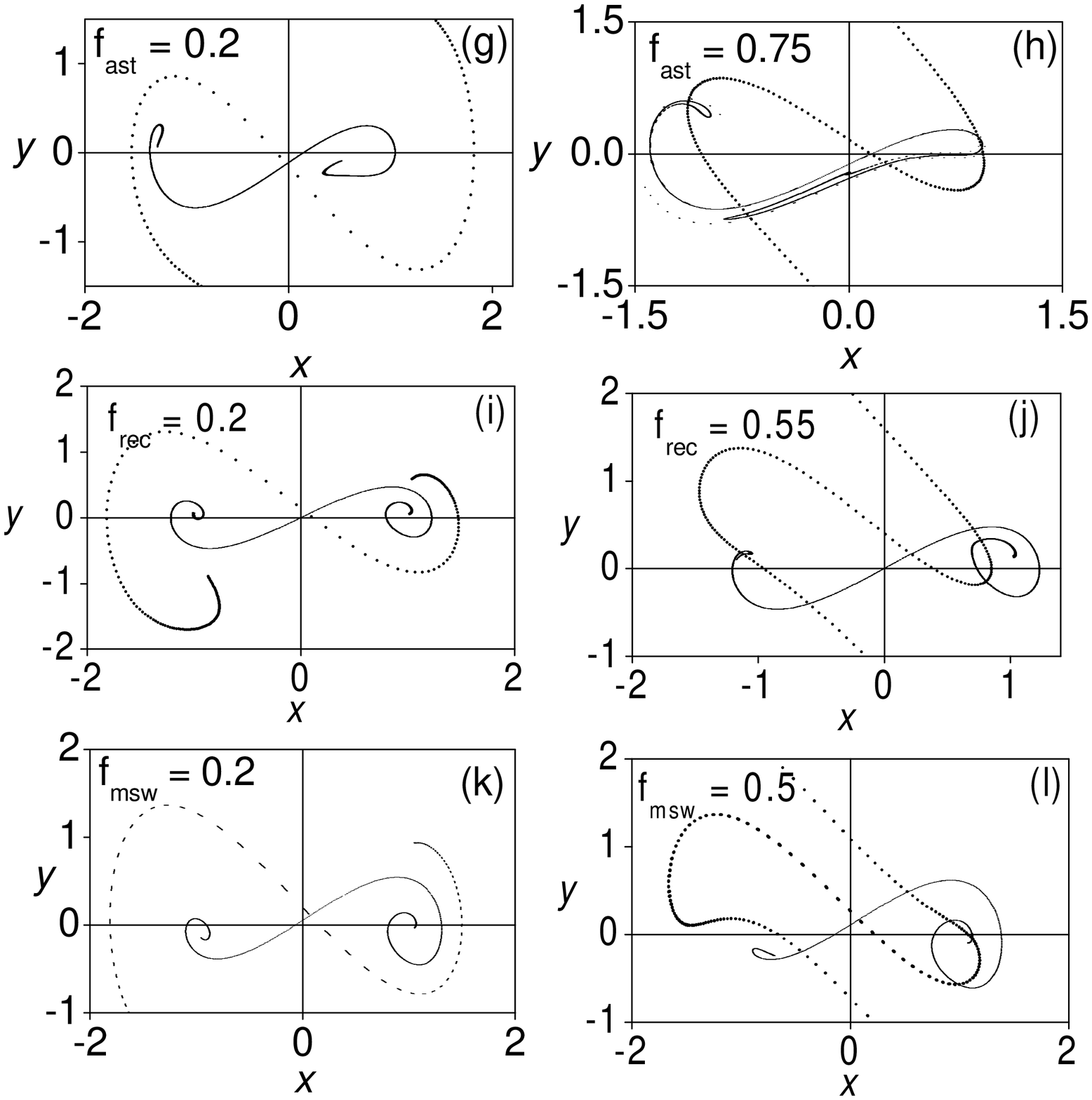}
\end{center}
\caption[]{continued...}\label{Fig32}
\end{figure}
\begin{figure}
\begin{center}
\includegraphics[width=0.4\columnwidth,clip]{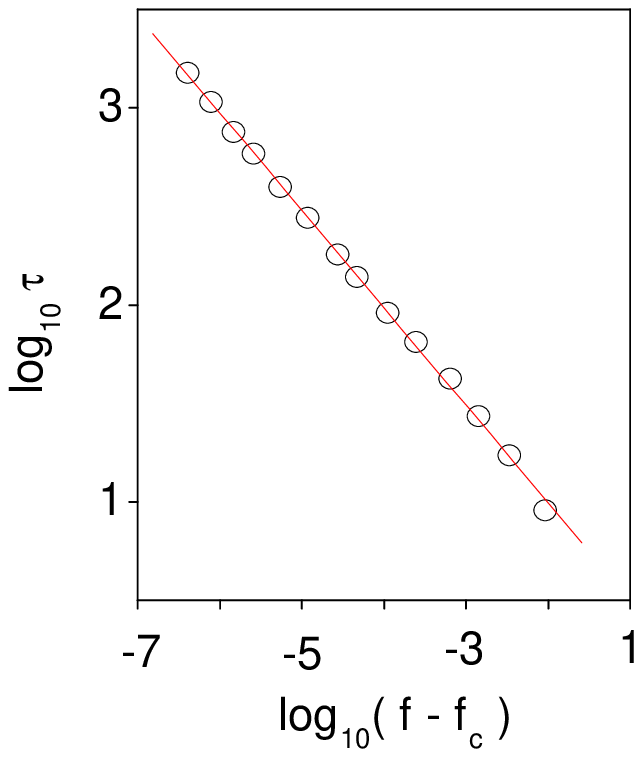}
\end{center}
\caption{Variation of $\tau$  as a function of $(f - f_{c})$.
$\tau$ is found to scale as 1.04166 $(f -
f_{c})^{-0.49}$.}\label{Fig4}
\end{figure}
\begin{figure}
\begin{center}
\includegraphics[width=0.4\columnwidth,clip]{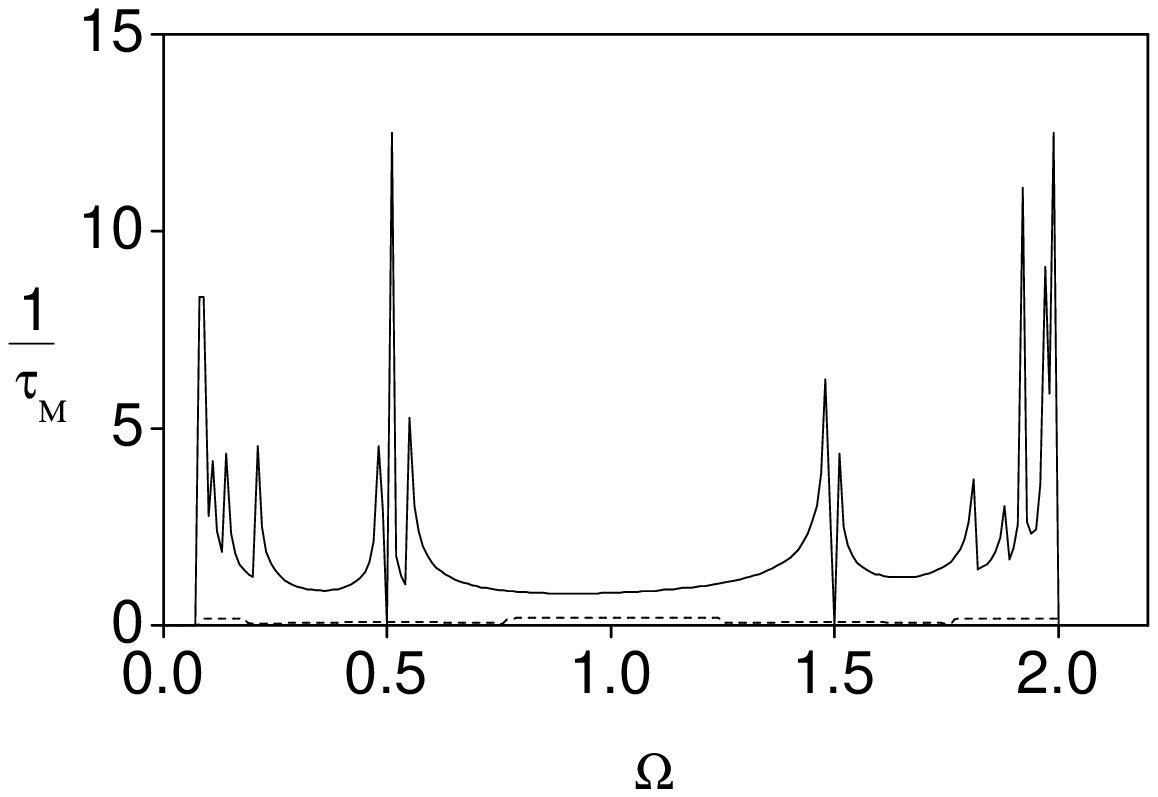}
\end{center}
\caption{$1/\tau_{M}$ versus $\Omega$ for the system (\ref{eq1}) with
$F(t)=f_{\sin} \sin \omega t + g_{\sin} \sin(\Omega t + \phi)$,  $g_{\sin} = 0.1$,
$f_{\sin} = 0.359$, $\alpha = 0.5$, $\beta = 1$,
$\omega_{0}^{2} = 1$, $\omega = 1$ and $\phi = 0$. Continuous
curve corresponds to positive sign while dashed curve corresponds
to negative sign in eq.(\ref{eq14}).}\label{Fig5}
\end{figure}
\begin{figure}
\begin{center}
\includegraphics[width=0.4\columnwidth,clip]{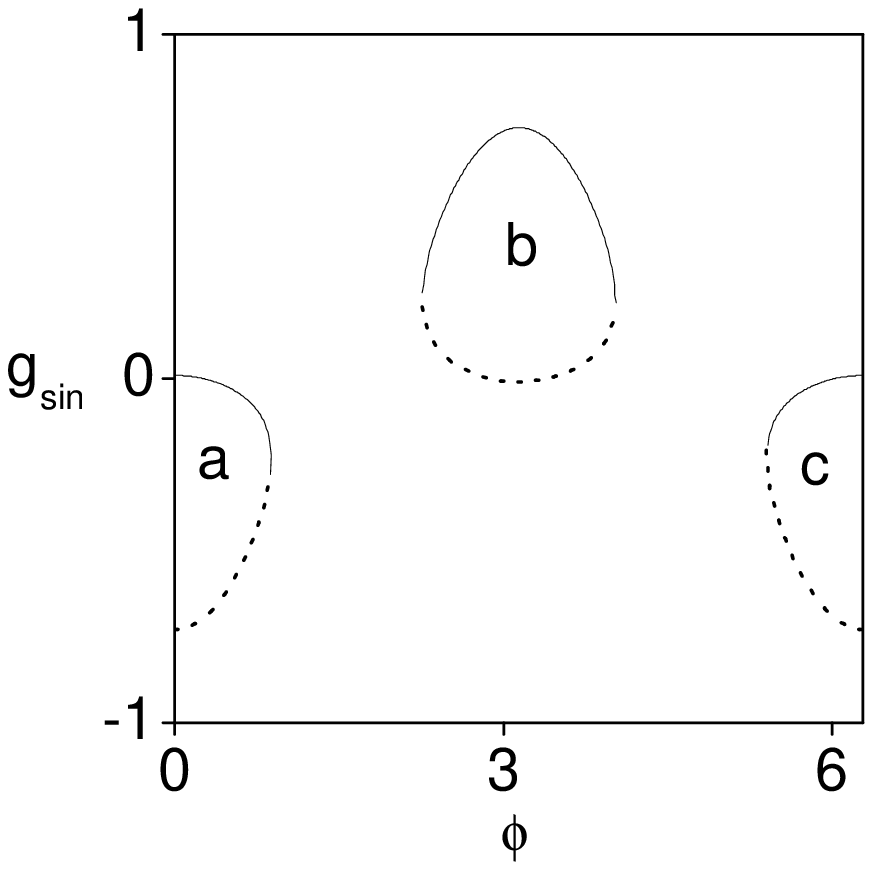}
\end{center}
\caption{Graph of the roots of $g_{\sin}^\pm$ of eq.(\ref{eq15}) with equality sign.
Horseshoe dynamics does not occur in the regions $\bf{a}$, $\bf{b}$, and
$\bf{c}$ enclosed by the curves. Continuous and dashed curves
represent $g_{\sin}^{+}$ and $g_{\sin}^{-}$ respectively. The
other parameters are $f_{\sin} = 0.359$, $\alpha= 0.5$, $\beta =
1$, $\omega = \Omega = 1$ and $\omega_{0}^{2} = 1$.}\label{Fig6}
\end{figure}
\begin{figure}
\begin{center}
\includegraphics[width=0.7\columnwidth,clip]{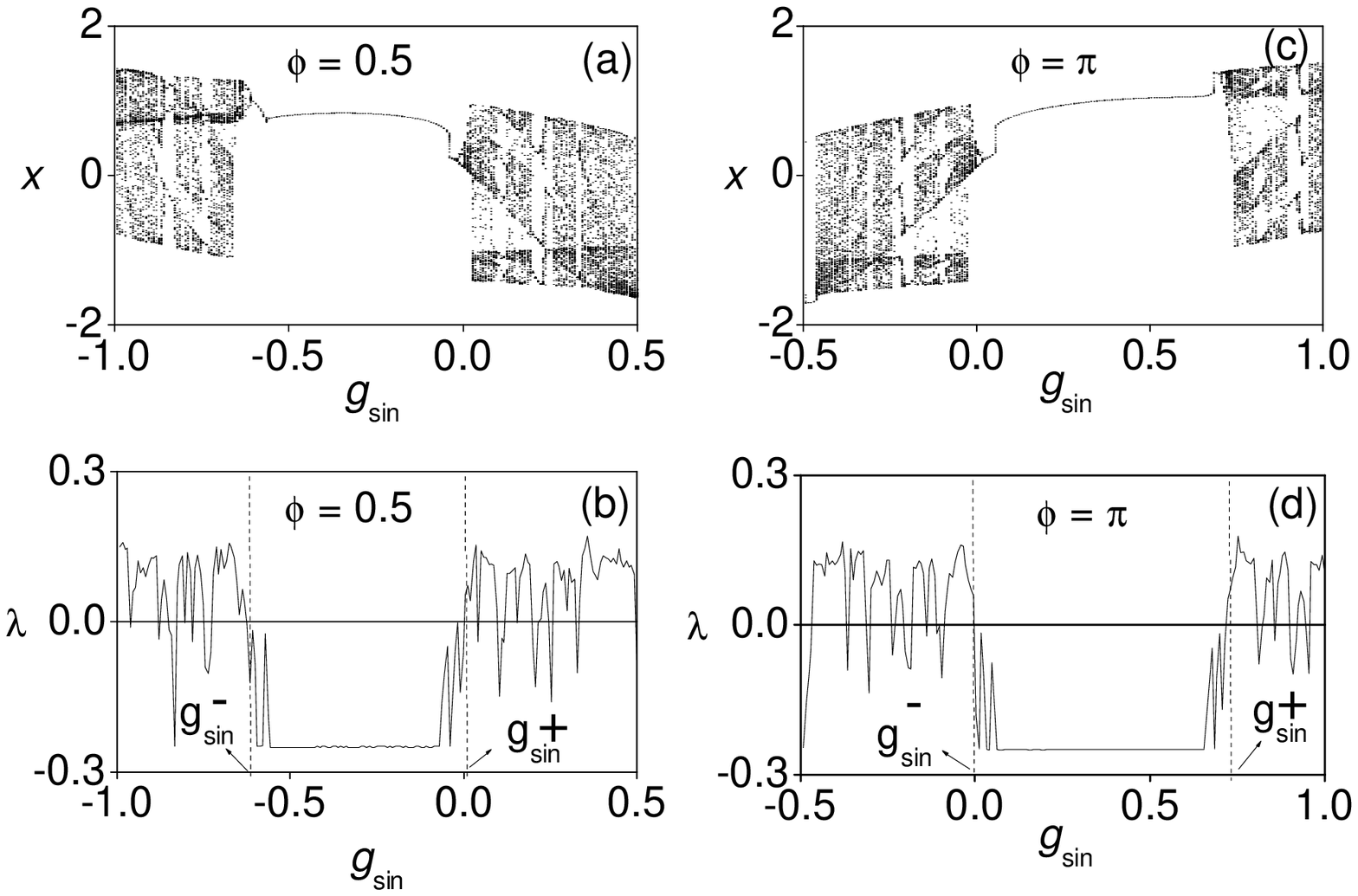}
\end{center}
\caption{Bifurcation diagrams and the corresponding maximal
Lyapunov exponent illustrating the effect of $\phi$ and $g$ on
chaotic dynamics. The parameters are fixed as $\alpha = 0.5$,
$\beta = 1$, $\omega_{0}^{2} = 1$, $\omega = \Omega = 1$,
$f_{\sin} = 0.359$. For $g_{\sin} = 0$ the system exhibits
chaotic motion.}\label{Fig7}
\end{figure}
\begin{figure}
\begin{center}
\includegraphics[width=0.4\columnwidth,clip]{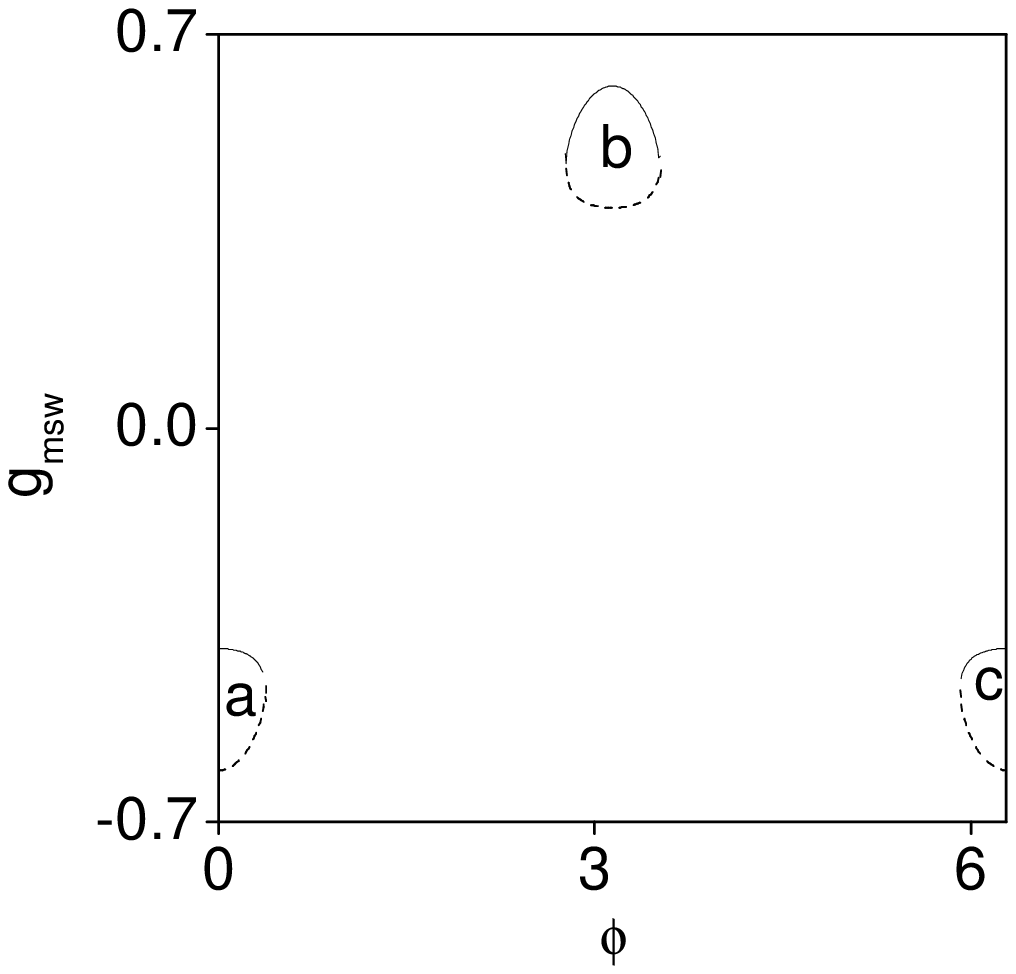}
\end{center}
\caption{Graph of $g_{\mbox{msw}}^{\pm}$.
Horseshoe dynamics does not occur in the regions $\bf{a}$,
$\bf{b}$, and $\bf{c}$ enclosed by the curves. 
The other parameters are $f_{\mbox{msw}} = 0.5$, $\alpha = 0.5$, $\beta =
1$, $\omega = \Omega = 1.0$ and $\omega_{0}^{2} = 1$.
}\label{Fig8}
\end{figure}
\begin{figure}
\begin{center}
\includegraphics[width=0.4\columnwidth,clip]{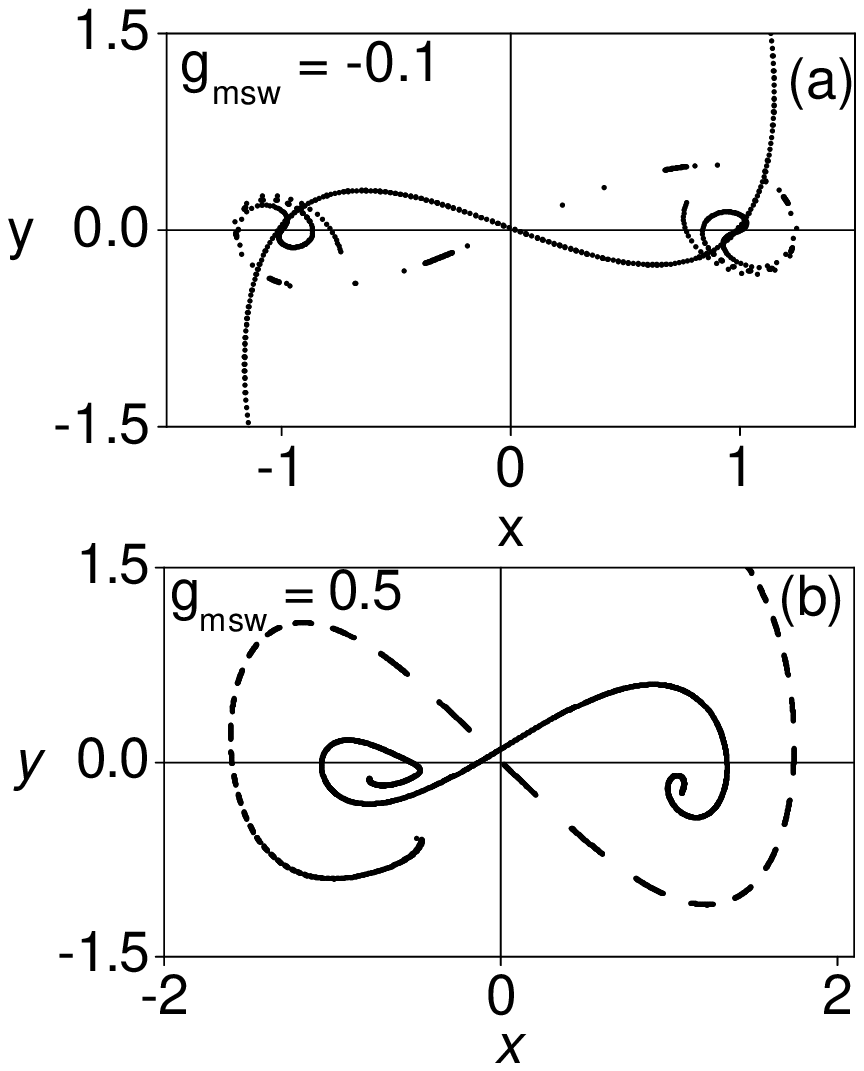}
\end{center}
\caption{Numerically computed part of perturbed homoclinic orbits
for (a) $g_{\mbox{msw}} = -0.1$ and (b) $g_{\mbox{msw}} = 0.5$.
The other parameters  are fixed at $\alpha = 0.5$, $\beta = 1$,
$\omega_{0}^{2} = 1$, $\phi = \pi$, $\Omega = \omega = 1$ and
$f_{\mbox{msw}} = 0.5$.
 }\label{Fig9}
\end{figure}
\end{document}